\documentclass[twocolumn]{aastex631}
\usepackage[utf8]{inputenc}
\usepackage[english]{babel}
\usepackage{cmap}
\usepackage{amsmath}
\usepackage{graphicx}
\usepackage{ amssymb }
\usepackage{aas_macros}
\hypersetup{pdfauthor=author}
\usepackage{xcolor}
\usepackage{comment}

\usepackage{natbib}
\usepackage{csquotes}

\begin{document}

\title{Towards constraining axions with polarimetric observations of the isolated neutron star RX J1856.5-3754}

\author{Aleksei Zhuravlev}
\affiliation{Faculty of Physics, Lomonosov Moscow State University,
Vorobjevy Gory 1,
Moscow, 119234, Russia}
\author{Roberto Taverna}
\affiliation{Dipartimento di Fisica e Astronomia, Universit\`a di Padova, via Marzolo 8, I-35131 Padova,
Italy}
\author{Roberto Turolla}
\affiliation{Dipartimento di Fisica e Astronomia, Universit\`a di Padova, via Marzolo 8, I-35131 Padova,
Italy}
\affiliation{MSSL-UCL, Holmbury St. Mary, Dorking, Surrey RH5 6NT, UK}
\email{zhuravlev.aa18@physics.msu.ru}

\begin{abstract}

Photon-axion mixing can create observable signatures in thermal spectra of isolated, cooling neutron stars. Their shape depends on the polarization properties of the radiation, which, in turn, are determined by the structure of stellar outermost layers. Here we investigate the effect of mixing on the spectrum and polarimetric observables, polarization fraction and polarization angle, using realistic models of surface emission. We focus on RX J1856.5-3754, the only source among the X-ray dim isolated neutron stars for which polarimetric measurements in the optical band were performed. Our results show that in the case of a condensed surface in both fixed and free-ion limits, the mixing can significantly limit the geometric configurations which reproduce the observed linear polarization fraction of 16.43\%. In the case of an atmosphere, the mixing does not create any noticeable signatures. Complementing our approach with the data from upcoming soft X-ray polarimetry missions will allow to obtain constraints on $g_{\gamma a} \sim 10^{-11}$ GeV$^{-1}$ and $m_a \lesssim 10^{-6}$ eV, improving the present experimental and astrophysical limits.

\end{abstract}

\section{Introduction}
\label{intro}

Since their first introduction in the late 1970s, axion-like particles (or simply axions) played a major role in modern extensions of the Standard Model of particle physics. Apart from explaining the absence of CP violation in strong interactions \citep{1977PhRvL..38.1440P, 1978PhRvL..40..279W}, they can provide a solution to various problems in physics and astrophysics (see reviews by \citealt{raffelt1996stars,2016PhR...643....1M,2020PhR...870....1D,2020arXiv201205029C}).

Primarily, axions manifest themselves through conversion into ordinary photons (and vice versa) in external magnetic fields. The strength of the interaction is mainly determined by the field geometry, coupling constant $g_{\gamma a}$, and axion mass $m_a$. Currently, a significant portion of the axion parameter space has been excluded by numerous experimental and astrophysical studies. Among laboratory-based searches, the more stringent bounds on the coupling constant over a wide axion mass range were obtained by the CERN Axion Solar Telescope (CAST), $g_{\gamma a} \leq 6.6 \times 10^{-11}$ GeV$^{-1}$ (95\% C.L., \citealt{2017NatPh..13..584A}). Planned missions such as IAXO \citep{2014JInst...9.5002A} promise a tremendous improvement over CAST, and have the potential of reaching down to $g_{\gamma a} \sim 10^{-12}$ GeV$^{-1}$. 

Various astrophysical searches typically attempt to find manifestations of axions in the radiation from sources observed through regions permeated by large-scale magnetic fields, such as in the case of galaxy clusters \citep{2016PhRvL.116p1101A,2017ApJ...847..101B,2018MNRAS.479.2243C,2020PhLB..80235252L}, blazars and quasars \citep{2011PhRvD..84l5019F,2019MNRAS.487..123G}, supernovae \citep{Payez:2014xsa}. Although in most cases the original spectrum of the source is not precisely known, the presence of mixing should result in the appearance of distinctive features. Their exact shape is determined by the choice of axion parameters and configuration of the external magnetic field. The possible constraints from such studies expand those of CAST, covering the region $g_{\gamma a} \sim 10^{-12}$ GeV$^{-1}$.

Another promising approach is to search for manifestations of axions in strongly magnetized compact objects, such as magnetic white dwarfs (mWDs) and neutron stars (NSs). First, axions may be produced thermally inside their cores with energies from several keV to several hundred keV through the nucleon or electron bremsstrahlung processes. Subsequently, they can escape the interior due to their feeble interaction with matter and turn into photons in the magnetic field surrounding the star, creating signatures in its hard X-ray spectrum. The absence of such features can be used to impose constraints on $\left( m_a, g_{\gamma a} \right)$, the neutron-axion coupling constant $g_{n a}$  \citep{raffelt1996stars,RAFFELT19901,2018JHEP...06..048F,2019PhRvD..99d3011S,2019JHEP...01..163F,2021PhRvD.103b3010L}, and the electron-axion coupling constant $g_{e a}$ \citep{2021arXiv210412772D}

In the case of white dwarfs, surface photons may undergo conversion as they propagate through the encompassing gaseous atmosphere \citep{2011PhRvD..84h5001G}. This could lead to an enhancement of the observed degree of linear polarization, which is quite low in most white dwarfs. In particular, for mWD PG 1015+014 the induced polarization degree exceeds the measured value of 5\% at $g_{\gamma a} \geq 2 \times 10^{-11}$ GeV$^{-1}$. Constraints from other mWDs with stronger magnetic fields can further expand this region, provided that the atmospheric plasma density is sufficiently low.

In the case of neutron stars, axions can create a number of distinctive features. First, their magnetospheres can be a site for the conversion of dark matter axions \citep{2009JETP..108..384P, 2020PhRvL.125q1301F}. Second, spectra of neutron stars typically include a thermal component, and surface photons may oscillate into axions as they propagate through the encompassing magnetosphere \citep{2006PhRvD..74l3003L,2012ApJ...748..116P}. Assuming the spectral energy distribution of an isotropic blackbody with uniform surface temperature and a dipolar magnetic field, \citealt{ZHURAVLEV2021136615} (hereafter Paper I) have shown that the conversion can reduce the optical O-mode flux by up to 30\% at the currently unconstrained values of $\left( m_a, g_{\gamma a} \right)$. The high-energy part of the spectrum remained unchanged in all cases. If the surface radiation of neutron stars includes a considerable fraction of O-mode photons, such manifestation of axions can create potentially observable effects for $g_{\gamma a} \sim 10^{-12}$ GeV$^{-1}$. 

The polarization properties of thermal emission are primarily determined by the structure of stellar outermost layers. It is commonly believed that the cooling surface is covered with a gaseous atmosphere that reprocesses the radiation coming from the underlying crustal layers \citep{2006RPPh...69.2631H, 2006MNRAS.373.1495V,2014PhyU...57..735P}. Its opacity for the X-mode photons is substantially reduced by the strong magnetic field, while that for the O-mode ones remains almost unaffected. Therefore, in this case, the mixing is unlikely to create detectable signatures, since the observed photons will be mostly polarized in the X-mode. On the other hand, a group of sources -- the so-called X-ray dim isolated neutron stars or XDINSs, also known as the ``Magnificent Seven" \cite[see e.g.][for a review]{2009ASSL..357..141T} -- may have a liquid or solid condensed surface due to a phase transition in the outermost layers induced by the strong magnetic field and relatively low temperature \citep{1997ApJ...491..270L,2001RvMP...73..629L,2007MNRAS.382.1833M,2010A&A...522A.111S}. As a result, no atmosphere will be present, and the polarization properties of the radiation will be mainly determined by the emissivity of the underlying metallic layer. Since the latter is of the same order in both X- and O-mode over a broad energy range, the emission of a condensed surface represents a promising place to search for manifestations of axions.

Among the ``Magnificent Seven", the best source for such study is RX J1856.5-3754 (RX J1856). It is the brightest and nearest of XDINSs, with $V=25.58$, a nearly $\lambda^{-4}$ optical-UV SED \citep{2001A&A...378..986V,2011ApJ...736..117K} and an X-ray spectrum well modeled by two blackbody components \citep{2012A&A...541A..66S}. The period $P \sim 7 \text{ s}$ and period derivative $\dot{P} \sim 3 \times 10^{-14} \text{ s s}^{-1}$ translate into a spindown magnetic field $B \sim 1.5 \times 10^{13}$ G \citep{2008ApJ...673L.163V}. 

More importantly, RX J1856 is the only XDINS for which polarimetric observations in the optical band were performed with the Very Large Telescope (VLT). \cite{2017MNRAS.465..492M} measured a phase-averaged polarization degree $\hat{\Pi}_{\mathrm{L}}=16.43 \% \pm 5.26 \%$ and a phase-averaged polarization position angle $\hat{\chi}_p= 145.39^\circ \pm 9.44^\circ$, computed east of the North Celestial Meridian. As shown by \citealt{2016MNRAS.459.3585G} (hereafter Paper II), polarimetric measurements can be effectively used to determine the structure of stellar outermost layers. The two models produce different polarization patterns at infinity -- in the case of an atmosphere, the radiation appears much more polarized than in the case of a condensed surface. As a result, by combining the optical measurements with the data from upcoming soft X-ray polarimetry missions, such as XPP \citep[\citealt{2019arXiv190710190J}, the follow-up mission of IXPE,][]{2016SPIE.9905E..17W}, it is potentially possible to unambiguously determine the origin of thermal emission from RX J1856.

The goal of our paper is to describe how the presence of photon-axion conversion affects the total flux and the polarization observables of a gaseous atmosphere and a condensed surface, focusing on the case of RX J1856. In Section 2, we provide an overview of the photon-axion interaction in strong magnetic fields and summarize the main properties of both models of surface emission. Section 3 describes our approach to calculating the polarization observables and the simplified method we adopt to account for the effect of photon-axion conversion. The results for RX J1856 are presented in Section 4. The observability of axion effects and methods for deriving constraints on their parameters using optical and X-ray polarimetry data are discussed  in Section 5.

\section{Theoretical background}
\label{theory}

Here we briefly summarize the basic theory behind our calculations; for more detailed explanations we refer to Papers I and II. We use Lorentz–Heaviside units with $\hbar = c = 1$ and $\alpha = e^2/4\pi \sim 1/137$.

\subsection{Photon-axion mixing}
\label{mixing}

The oscillations of an O-mode photon with energy $\omega$ and an axion in the external magnetic field are described by the following system: 

\begin{align}\label{eqn_2_matrix_form}
    \left(\omega + \begin{pmatrix}
    \Delta_\parallel + \Delta_p & \Delta_{M}\\
    \Delta_{M}    & \Delta_{a}
    \end{pmatrix} - i\partial_z \right)
    \begin{pmatrix}
    A_\parallel \\ a
    \end{pmatrix} = 0,
\end{align}

\begin{equation}
\begin{aligned}
\Delta_{a} &= -\frac{m_a^2}{2\omega}, & \Delta_{M} &= \frac{1}{2} g_{\gamma a} B \sin\Theta, \\
\Delta_{p} &= -\frac{\omega_p^2}{2\omega}, & \Delta_{\parallel} &= \frac{1}{2} \omega\left( n_{\parallel} - 1 \right),
\end{aligned}
\end{equation}

\noindent where $A_\parallel$ and $a$ denote the amplitudes of the ordinary photon state and the axion, respectively; $\omega_{p}^{2}=4 \pi \alpha n_{e} / m_{e}$ is the plasma frequency squared, the $z-$ axis is along the direction of photon propagation, and $\Theta$ is the angle between the latter and the direction of the local magnetic field \citep{1988PhRvD..37.1237R}. We assume that the electron density is of the order of the Goldreich-Julian value, $n_{e} =  -\boldsymbol{\Omega} \cdot \boldsymbol{B} / e$, where $\boldsymbol{\Omega}$ is the stellar angular velocity \citep{1969ApJ...157..869G}. The photon refractive index $n_{\parallel}$ deviates from  unity due to the presence of a strong magnetic field \cite[see e.g.][]{2006RPPh...69.2631H}. 

As shown in Paper I, for the currently allowed values of the coupling constant $g_{\gamma a}$ only the first half-period of oscillations can be present in the magnetosphere of a neutron star (weak-oscillation mode), i.e. photons convert into axions and do not convert back. This is due to the fact that either the mixing term is dominated by the QED term $\left( \Delta_{M} \ll \Delta_{\parallel} \right)$, or the oscillation length becomes much larger than the scale length of the stellar magnetic field. In addition, only the low-energy radiation can undergo conversion, while the X-ray flux always remains unchanged. For $B_0 = 10^{13}$ G, $g_{\gamma a} = 2 \times 10^{-11}$ GeV$^{-1}$, and $m_a = 10^{-8}$ eV, 30\% of the optical O-mode photons (redshifted energy $\omega_\infty = 1 \text{ eV}$) do not reach the observer and instead turn into axions. In the following, we express the resulting change in the spectrum in terms of the differential modification factor $\Lambda$, which is defined as the relative fraction of the initial photon flux with a given energy, reaching the observer without conversion. In particular, a 30\% decrease in the optical radiation corresponds to $\Lambda = 0.7$.

\subsection{Intrinsic polarization degree}
\label{intr_poldeg}

The cooling surface of a neutron star is commonly modeled by a gaseous atmosphere in radiative and hydrostatic equilibrium \cite[see e.g.][for a review]{2014PhyU...57..735P}. A quite general property is that the opacity for X-mode photons below the electron cyclotron frequency is substantially reduced by the strong magnetic field, while that for the O-mode ones remains almost unchanged \cite[see][for a comprehensive overview of the physics in strong magnetic fields]{2006RPPh...69.2631H}. As a result, the intensity of the X-mode photons becomes much larger than that of the O-mode ones, and the emergent radiation appears highly polarized. Here we restrict to the case of a fully ionized pure H atmosphere and solve the radiative transfer equation using the numerical method developed by \citet[][see also \citealt{Lloyd_2003,2006MNRAS.366..727Z}]{2003astro.ph..3561L}. The code has four input parameters: the local magnetic field strength, the effective temperature, the angle between the local magnetic field and the surface normal, and the surface gravity. To compute the emergent intensity for a range of photon energies and emission directions, a complete linearization technique for the two normal polarization modes is employed in a plane-parallel slab, see Paper II. 

The polarization pattern produced by a particular surface emission model can be conveniently expressed in terms of the intrinsic polarization degree: 

\begin{equation}\label{eq:intrinsic}
\Pi_\text{L}^\text{EM} = \frac{F_\text{X} - F_\text{O}}{F_\text{X} + F_\text{O}},
\end{equation}

\noindent where $F_\text{X,O}$ is the monochromatic, phase-averaged flux in each mode, obtained by integrating the intensity over the visible part of the star surface. By definition, $\Pi_\text{L}^\text{EM} = 1$ for the radiation 100\% initially polarized in the X-mode, $\Pi_\text{L}^\text{EM} = -1$ for that in the O-mode, and $\Pi_\text{L}^\text{EM} = 0$ for the unpolarized emission. As shown in Paper II, the intrinsic polarization degree of an atmosphere is $\Pi_\text{L}^\text{EM} \approx 0.87$ in the optical band and $\Pi_\text{L}^\text{EM} \approx 0.99$ in the X–ray band for all viewing geometries, so the radiation is dominated by the X-mode photons. 

On the other hand, it has been suggested that at  the low surface temperature $\left( \lesssim 100 \text{ eV} \right)$ and  strong magnetic field $\left( \gtrsim 10^{13} \text{ G} \right)$, typical of XDINSs, a phase transition between a gaseous and a condensed state may occur. As a result, the underlying metallic layer of the stellar crust will be exposed, and the emergent spectrum will be determined by its radiative properties \citep{1997ApJ...491..270L, 2001RvMP...73..629L,2004ApJ...603..265T, 2007MNRAS.382.1833M, 2009ASSL..357..141T, 2014PhyU...57..735P}. Two limiting cases are typically considered: free-ions, in which the effects of the lattice on the interaction of the electromagnetic waves with ions are neglected, and fixed-ions, in which the lattice interaction with electromagnetic waves dominates the ion response. The true spectral properties are expected to lie between these two limits. 

Following Paper II, we adopt the analytic descriptions developed by \cite{2012A&A...546A.121P} to calculate the intensity in two modes. The reflectivity is first obtained by applying Snell's law at the interface between the vacuum and the condensed phase. The emissivity  $j_{\omega}$ is related to the latter via Kirchhoff's law, and the intensity is 

\begin{equation}
\begin{gathered}
I_{\omega, \mathrm{O}}=j_{\omega, \mathrm{O}} B_{\omega}(T), \\
I_{\omega, \mathrm{X}}=j_{\omega, \mathrm{X}} B_{\omega}(T),
\end{gathered}
\end{equation}

\noindent where $B_{\omega}(T)=\omega^{3} /\left[4\pi^3\left( \exp\left( \omega / T \right)-1 \right)\right]$ is the blackbody radiance. 

Contrary to the atmospheric model, the emissivity of a condensed surface in the X- and O-mode is of the same order over a broad energy range \citep{2010A&A...522A.111S}. In the optical band, $\Pi_\text{L}^\text{EM}$ can reach $-0.3$ in the free-ion and $-0.5$ in the fixed-ion limits for the most favorable viewing geometries, with the emission being polarized mostly in the O-mode. In the X-ray band, the radiation is almost unpolarized, $|\Pi_\text{L}^\text{EM}| \lesssim 0.07$ in both limits.

\section{The model}
\label{model}

To calculate the effects of photon-axion conversion using realistic models of  surface emission, we combine the methods presented in Papers I and II. The long computation time implied by such a direct approach can be significantly reduced by making a number of simplifications which are discussed below.

We restrict to the case in which the stellar magnetic field is dipolar,

\begin{equation}
    \boldsymbol{B}^{\text{dip}} = \frac{B_p}{2} \left(\frac{R_{\rm NS}}{r}\right)^3 \begin{pmatrix}
    2 f_{\text{dip}} \cos\theta \\
    g_{\text{dip}} \sin\theta \\
    0
    \end{pmatrix},
\end{equation}

\noindent where $B_p$ is the polar field strength, $R_{\rm NS}$ is the stellar radius, and $r, \theta$ are the radial coordinate and the magnetic colatitude, respectively. The functions $f_{\text {dip }}$ and $g_{\mathrm{dip}}$ account for the relativistic corrections \citep{1986SvA....30..567M, 1996ApJ...473.1067P}. The polar magnetic field is taken to be $B_p=10^{13}$ G, which is compatible with the spin-down estimate for RX J1856 \citep{2008ApJ...673L.163V}. The spin period is the measured one, $P = 7$ s \citep{2007ApJ...657L.101T}, while the stellar radius and mass are taken as  $R_{\rm NS} = 12$ km, $M_{\rm NS}=1.5 M_\odot$, respectively. The surface temperature distribution is that induced by a core-centered dipole, see Paper II for details.

\subsection{Ray tracing}
\label{ray_tracing}

Thermal emission of a cooling neutron star is calculated using the ray tracing method presented in \citealt{2006MNRAS.366..727Z} (see also \citealt{2015MNRAS.454.3254T} and Paper II). First, the source geometry is determined by choosing the angle $\chi$ between the line of sight $\boldsymbol{\ell}$ and the spin axis $\mathbf{p}$, as well as the angle $\xi$ between the magnetic (dipole) axis $\mathbf{b}_{\mathrm{dip}}$ and the spin axis. Then, the monochromatic flux detected by a distant observer is obtained by summing the contribution of each surface element which is in view at a given rotational phase, taking into account the effects of ray bending,

\begin{equation}
F_{\nu}(\gamma)=\left(1-\frac{R_{\mathrm{s}}}{R_{\mathrm{NS}}}\right) \frac{R_{\mathrm{NS}}^{2}}{D^{2}} \int_{0}^{2 \pi} d \Phi_{\mathrm{S}} \int_{0}^{1} I_{\nu}(\mathbf{k}, \theta, \phi) d u^{2}.
\end{equation}

\noindent Here $I_{\nu}$ is the specific intensity, which in general depends on the photon frequency $\nu$, emission direction $\mathbf{k}$, and position on the stellar surface, $R_{\rm s}$ is the Schwarzschild radius, $D$ the source distance and $x=R_{\rm s}/R_{\rm NS}$. The intensity is naturally written in terms of the polar angles $(\theta, \phi)$ of a co-rotating coordinate system $(x, y, z)$, with the $z$-axis parallel to $\mathbf{b}_{\text{dip}}$ and the $x$-axis orthogonal to both $\mathbf{p}$ and $\mathbf{b}_{\text{dip}}$. However, the integration is performed in a fixed reference frame $(X, Y, Z)$, with the $Z$-axis along the line of sight and the $X$-axis in the $(\boldsymbol{\ell}, \mathbf{p})$ plane, with the associated polar angles $\left(\Theta_{\mathrm{s}}, \Phi_{\mathrm{S}}\right)$.  The transformations linking $\left(\Theta_{\mathrm{s}}, \Phi_{\mathrm{S}}\right)$ and $(\theta, \phi)$ are given in Paper II. Finally, $u=\sin \bar{\Theta}$, and the angles $\Theta_{\mathrm{S}}$ and $\bar{\Theta}$ are related by the ``ray tracing'' integral

\begin{equation}
\bar{\Theta}=\int_{0}^{1 / 2} \frac{d v \sin \Theta_{s}}{\left[(1-x) / 4-(1-2 v x) v^{2} \sin ^{2} \Theta_{s}\right]^{1 / 2}}.
\end{equation}

\subsection{Polarization radius}
\label{pol_radius}

According to QED, the presence of a strong magnetic field makes the vacuum around a neutron star anisotropic. The dielectric and magnetic permeability tensors are modified by virtual electron-positron pairs, which impacts on the polarization properties of radiation. As a linearly polarized electromagnetic wave propagates in the magnetized vacuum near the star, its electric field can instantly follow the direction of the external magnetic field. Up to the adiabatic (polarization-limiting) radius,

\begin{equation}
r_{p} \simeq 7.9\left(\frac{B_p}{10^{13} \text{ G}}\right)^{2 / 5}\left(\frac{\omega_\infty}{1 \text{ eV}}\right)^{1 / 5}\left(\frac{R_{\rm NS}}{12\,{\rm km}}\right)^{1/5} R_{\mathrm{NS}},
\end{equation}

\noindent the photons emitted at the surface maintain their initial polarization state \citep{2002PhRvD..66b3002H,2003MNRAS.342..134H,2015MNRAS.454.3254T}. Around $r_p$, the coupling weakens and the photon electric field is no longer able to properly follow the variation of the local magnetic field; at $r \gg r_p$, its direction freezes. Using the adiabatic radius approximation, it is possible to take into account the effects of vacuum polarization without numerically integrating the wave equations along each photon trajectory. As shown in \cite{2015MNRAS.454.3254T}, one can assume mode locking up to $r_{p}$, and keep the direction of the wave electric field constant at $r>r_p$. As a result, the polarization properties will be determined by the direction of the local magnetic field at $r_p$, as well as by the intrinsic polarization degree.

To obtain the polarization observables as detected by a distant instrument, the Stokes parameters of individual photons at $r_p$ need to be rotated so that they refer to the same coordinate system \cite[see][for details]{2015MNRAS.454.3254T}. The collective Stokes parameters $I, Q, U$ are then simply the sum of the individual parameters, and the observed polarization fraction and polarization angle are given by

\begin{align}
\Pi_{\mathrm{L}}&=\frac{\sqrt{Q^{2}+U^{2}}}{I}\\
\chi_p&=\frac{1}{2} \arctan \left(\frac{U}{Q}\right)
\end{align}

\subsection{Conversion radius}
\label{conv_radius}

The main effects of the photon-axion interaction can be taken into account using the conversion-radius approximation. This method is based on several simplifications which are discussed below.

First, we utilize the fact that the relative converted fraction of the O-mode flux varies by no more than 2\% between different photon trajectories due to the weak-oscillation mode, which is present at all allowed values of $g_{\gamma a}$. Having fixed the magnetic field geometry, we use the code of Paper I to numerically integrate Equations (\ref{eqn_2_matrix_form}) over a large number of ray paths, taking into account the ordinary photons only, and obtain the average converted fraction $\bar{\Lambda}$. Furthermore, we assume that the mixing occurs instantly at a specific point, rather than gradually over a magnetospheric region. We introduce the conversion radius $r_c$, defined as the distance from the stellar surface within which 80\% of the effect takes place; for the optical radiation of RX J1856, $r_c \sim 200$--$300$ km. As a result, to account for the presence of axions we can simply multiply the O-mode flux of each photon ray by $\bar{\Lambda}$ at the conversion radius, instead of performing the time-consuming integration.

The adiabatic-radius method for taking into account the vacuum polarization is very similar to the conversion-radius approximation. In fact, these two steps can be combined. The dipolar magnetic field does not change its direction significantly between $r_p$ and $r_c \sim \left( 2-3 \right) r_p$; due to the weak-oscillation mode, the slight depolarization in this region may be safely neglected when calculating the effect of conversion. As a result, we can simply assume that $r_c = r_p$, and simultaneously decrease the O-mode flux and compute the Stokes parameters of the remaining photons at the adiabatic radius. By using these simplifications, we can fully utilize the robustness of the ray-tracing code and avoid solving the photon-axion propagation equations, which significantly reduces the computational time.

\section{Results}
\label{results}

\begin{figure}
\centering
\includegraphics[width=0.9\linewidth]{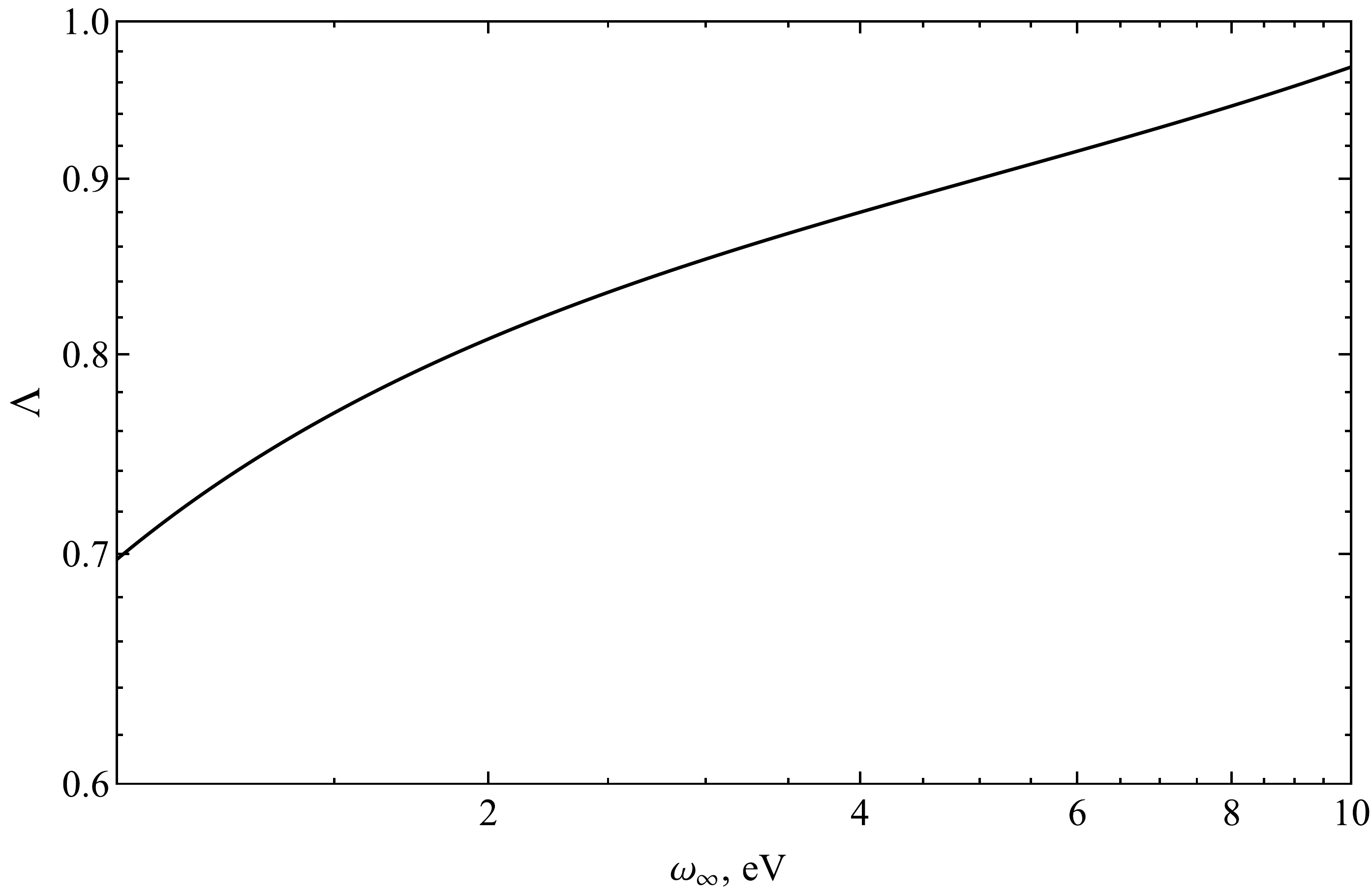}
\caption{Average modification factor $\bar{\Lambda}$ for $g_{\gamma a} = 2 \times 10^{-11}$ GeV$^{-1}$ and $m_a = 10^{-8}$ eV, as a function of the photon energy. Here the magnetic field geometry is an aligned dipole with $B_p = 10^{13}$ G and $\theta = 60^\circ$.
\label{fig:fig0}}
\end{figure}

In this section we present our simulations of the photon-axion interaction in the emission from a condensed surface in free/fixed ion limits, as well as a magnetized hydrogen atmosphere. We typically use the axion parameters $g_{\gamma a} = 2 \times 10^{-11}$ GeV$^{-1}$ and $m_a = 10^{-8}$ eV, which are below the present experimental constraints obtained by CAST \citep{2017NatPh..13..584A}. The corresponding average modification factor is shown in Figure \ref{fig:fig0}.

We make specific reference to RX J1856, for which \cite{2017MNRAS.465..492M} measured a phase-averaged polarization degree $\hat{\Pi}_{\mathrm{L}}=16.43 \% \pm 5.26 \%$ and a phase-averaged polarization position angle $\hat{\chi}_p= 145.39^\circ \pm 9.44^\circ$ in the VLT-FORS2 $\nu_{\rm HIGH}$ filter ($2.0-2.5$ eV). They were able to derive constraints on the source geometry by comparing the observed values of $\Pi_{\mathrm{L}}$ and of the X-ray pulsed fraction with the ones predicted by different surface emission models, as functions of the two angles $\xi$ and $\chi$ (see Section \ref{intr_poldeg}). In particular, it is either $\left( \xi,\chi \right) \sim \left( 5^\circ, 21^\circ \right)$ or $\left( 3^\circ, 52^\circ \right)$ for a fixed-ion condensed surface and $\left( \xi,\chi \right) \sim \left( 14^\circ, 3^\circ \right)$ for an atmosphere.

\begin{figure}
\centering
\includegraphics[width=0.9\linewidth]{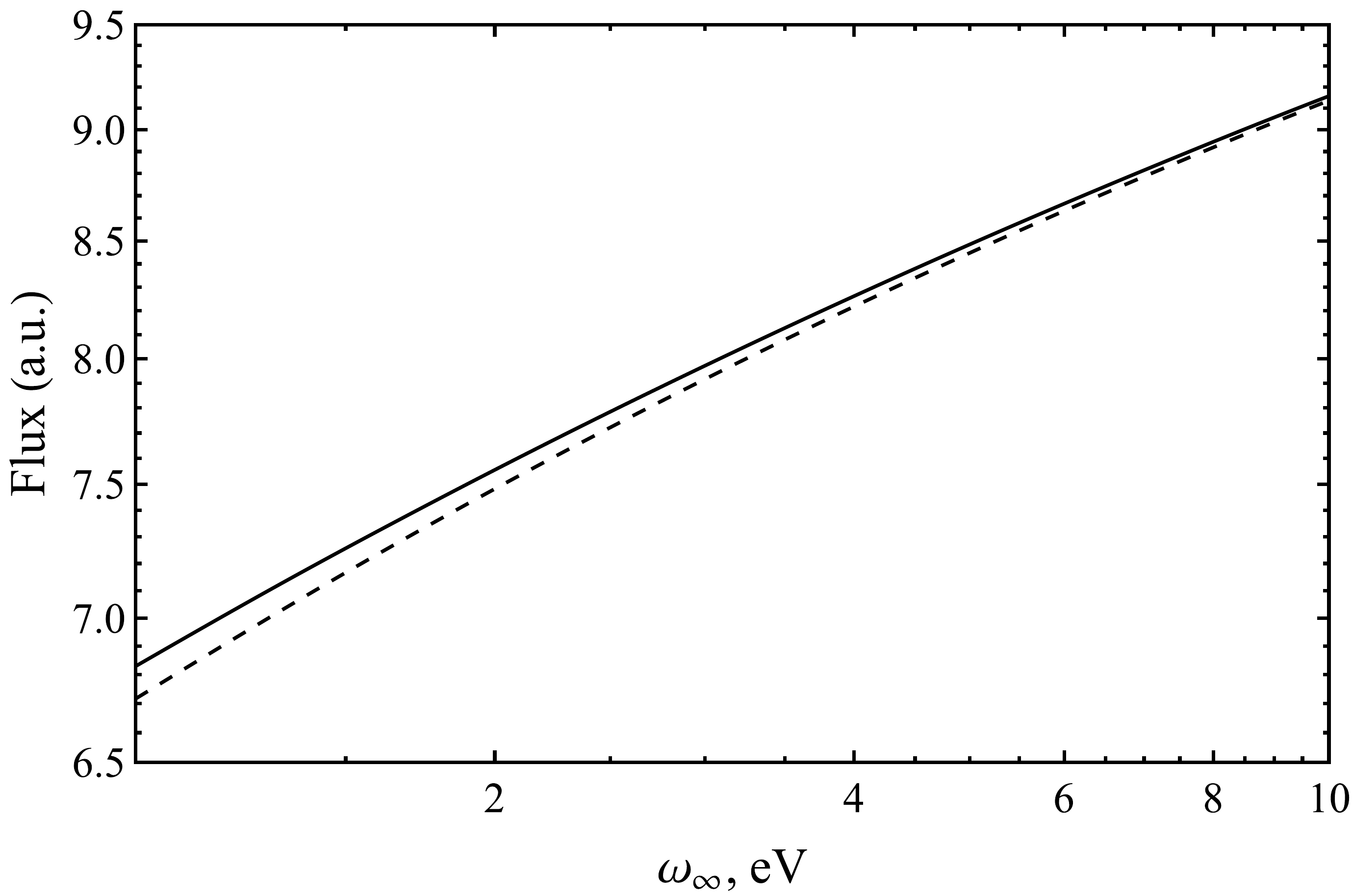}
\caption{Number flux (in arbitrary units) for the condensed surface in the fixed-ion limit as a function of the photon energy: without (solid line) and with mixing (dashed line). The source geometry is $\left( \xi,\chi \right) = \left( 5^\circ, 21^\circ \right)$ and the remaining parameters are those of RX J1856. 
\label{fig:fig1}}
\end{figure}

We first consider the case of a condensed surface in the fixed-ion limit. Figure \ref{fig:fig1} illustrates the optical spectra with and without conversion for $\left( \xi,\chi \right) = \left( 5^\circ, 21^\circ \right)$; the total flux in the former case turns out to be reduced by around 15\% with respect to that computed in the absence of mixing. This is expected since the intrinsic polarization degree is negative, and the radiation is dominated by the O-mode photons for which conversion may occur (see Paper II). However, the effect may be difficult to distinguish observationally. The optical counterparts of XDINSs are faint, and there are considerable uncertainties in their spectral measurements. In particular, for RX J1856 the change in total intensity by 15\% lies well within the errors  \cite[][]{2001A&A...378..986V}. As a result, polarimetric observables should be our primary tool in detecting the presence of axions, while the spectra are best used to validate our calculations. 

\begin{figure*}
\includegraphics[width=6.cm]{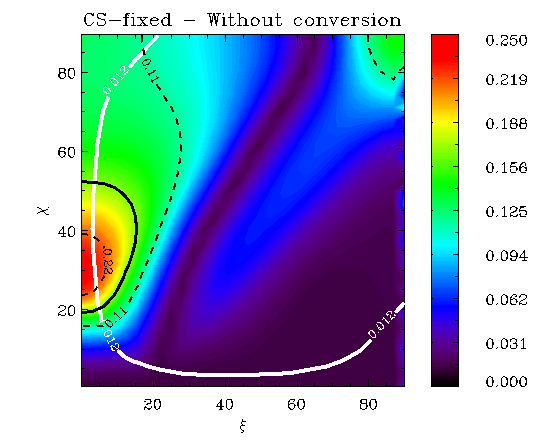}
\includegraphics[width=6.cm]{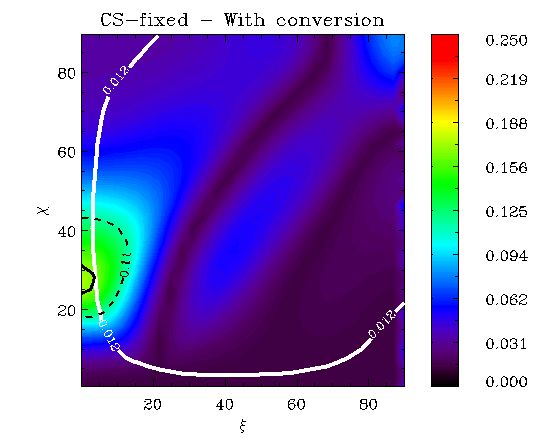}
\includegraphics[width=6.cm]{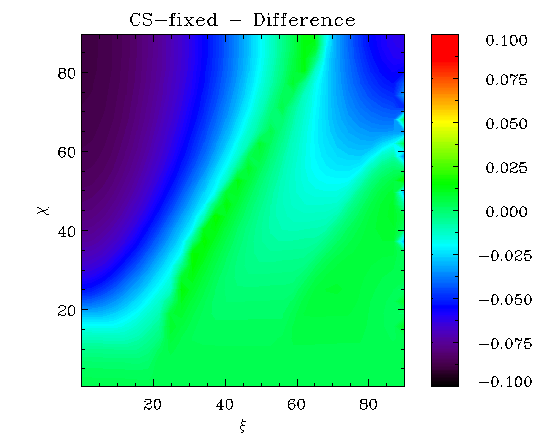}
\caption{Contour plots of the phase-averaged polarization fraction in the VLT-FORS2 $\nu_{\rm HIGH}$ filter ($2.0-2.5$ eV) for the condensed surface in fixed-ion limit. The solid black line indicates the measured value of $\bar{\Pi}_{\mathrm{L}} = 16.43\%$, while the dashed black lines mark the $1\sigma$ errors. The white line shows the combinations of $\left( \xi, \chi \right)$ which reproduce the observed X-ray pulsed fraction of 1.3\% \citep{2007ApJ...657L.101T}.
\label{fig:fig2}}
\end{figure*}

Figure \ref{fig:fig2} illustrates the phase-averaged polarization fraction with and without conversion, as well as the difference between the two  cases. As we can see, the predicted $\Pi_{\mathrm{L}}$ can indeed match the value measured in RX J1856 for some geometrical configurations in the absence of mixing. However, a 20\% decrease in the O-mode flux significantly reduces the overall polarization, so that $\hat{\Pi}_{\mathrm{L}}$ is barely reached if mixing occurs, and a much smaller portion of the $\xi-\chi$ plane lies within the $1\sigma$ error region. This result can be interpreted as follows. Without conversion, the amount of O-mode photons exceeds that of the X-mode ones, leading to a moderate degree of polarization. When the conversion is taken into account, the fluxes in the two modes become closer to each other, making radiation less polarized. According to our calculations, for lower values of $\bar{\Lambda}$ the radiation becomes dominated by X-mode photons, and the observed $\Pi_{\mathrm{L}}$ starts increasing. Therefore, our main conclusion is that for allowed values of $\left( m_a, g_{\gamma a} \right)$, the corresponding slight decrease in the O-mode flux can significantly reduce the polarization fraction.

Since the measured $\hat{\Pi}_{\mathrm{L}} = 16.43\%$ can still be reached at some combinations of $\left( \xi, \chi \right)$ despite the presence of conversion, we can additionally calculate the pulsed fraction induced by a given magnetic field geometry and compare it with the observational data. For RX J1856, its value is very low (1.3\% in the X-ray band, \citealt{2007ApJ...657L.101T}), which implies that either $\xi$ or $\chi$ is small $\left( \lesssim 15^\circ \right)$. As pointed out in \cite{2017MNRAS.465..492M}, this method allows to impose a constraint on the magnetic field geometry and thus exclude parts of its parameter space. Also, note that the region with $\xi \geq \chi$ is characterized by a vanishingly low degree of polarization, regardless of the intrinsic one at the source. The reason for that is the frame rotation of the Stokes parameters (see \citealt{2015MNRAS.454.3254T} and Paper II), which prevents us from distinguishing the presence of conversion in this region. While not relevant for RX J1856 since the measured optical polarization is non-negligible, it may become an issue for other sources.

\begin{figure*}
\centering
\includegraphics[width=7cm]{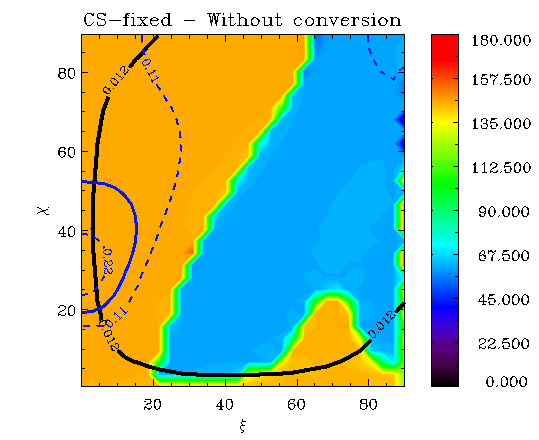}
\includegraphics[width=7cm]{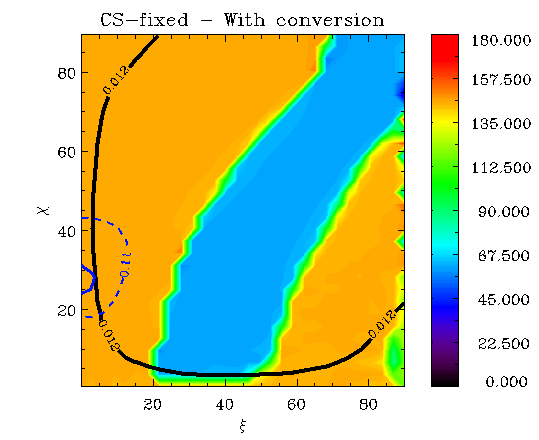}
\caption{Contour plots of the phase-averaged polarization angle in the VLT FORS2 $\nu_{\rm HIGH}$ filter for the condensed surface in fixed-ion limit. The black line indicates the magnetic field geometries for which the measured X-ray pulsed fraction is reproduced, while the blue lines mark those for $\hat{\Pi}_{\rm L}$. The polarimeter frame was rotated by an angle $\hat{\chi}_p = 145.39^\circ$ with respect to the fixed frame for consistency with the observations \citep{2017MNRAS.465..492M}.}
\label{fig:fig3}
\end{figure*}

Our calculations of the phase-averaged polarization angle are illustrated in Figure \ref{fig:fig3}. Here the polarimeter frame was rotated around the line-of-sight by an angle $\beta=\hat{\chi}_p$ with respect to the fixed frame $(X,Y,Z)$. As shown in \cite{2015MNRAS.454.3254T}, for the radiation mostly polarized in the O-mode $\chi_p$ becomes equal to the angle $\Psi$ between the polarimeter reference axis and the $X$-axis, and to $\Psi+90^\circ$ for that in the X-mode. Since the ordinary photons dominate in the region of $\xi-\chi$ plane constrained by the measured $\hat{\Pi}_{\rm L}$ and X-ray pulsed fraction, we need to rotate the polarimeter frame by $\hat{\chi}_p = 145.39^\circ$ in order to ensure consistency with the observations. 

As it can be seen, the value of $\chi_p$ slightly changes only in the region with $\xi \geq \chi$. This is due to the fact that the polarization angle reflects the global direction of the photon electric field, which, in turn, depends on the direction of the magnetic field at the adiabatic radius (see Paper II). As a result, the observed $\chi_p$ should reflect the “phase-averaged” direction of the magnetic field at $r_a$, and the jumps by $90^{\circ}$ arise from an even slight predominance of ordinary photons over extraordinary ones or conversely. Using the intrinsic polarization degree and our calculations of the intensity, we can infer that the O-mode flux exceeds that of the X-mode for most magnetic field geometries, both with and without conversion. As a result, the mixing changes the value of $\chi_p$ only in the region with a vanishingly small degree of polarization, where even a single additional photon can modify the direction of the phase-averaged wave electric field. Since $\chi_p$ remains constant for the source geometries which reproduce the measured $\Pi_{\rm L}$, we cannot utilize the measurement of polarization angle in the same way as that of the polarization fraction.

\begin{figure}
\centering
\includegraphics[width=0.9\linewidth]{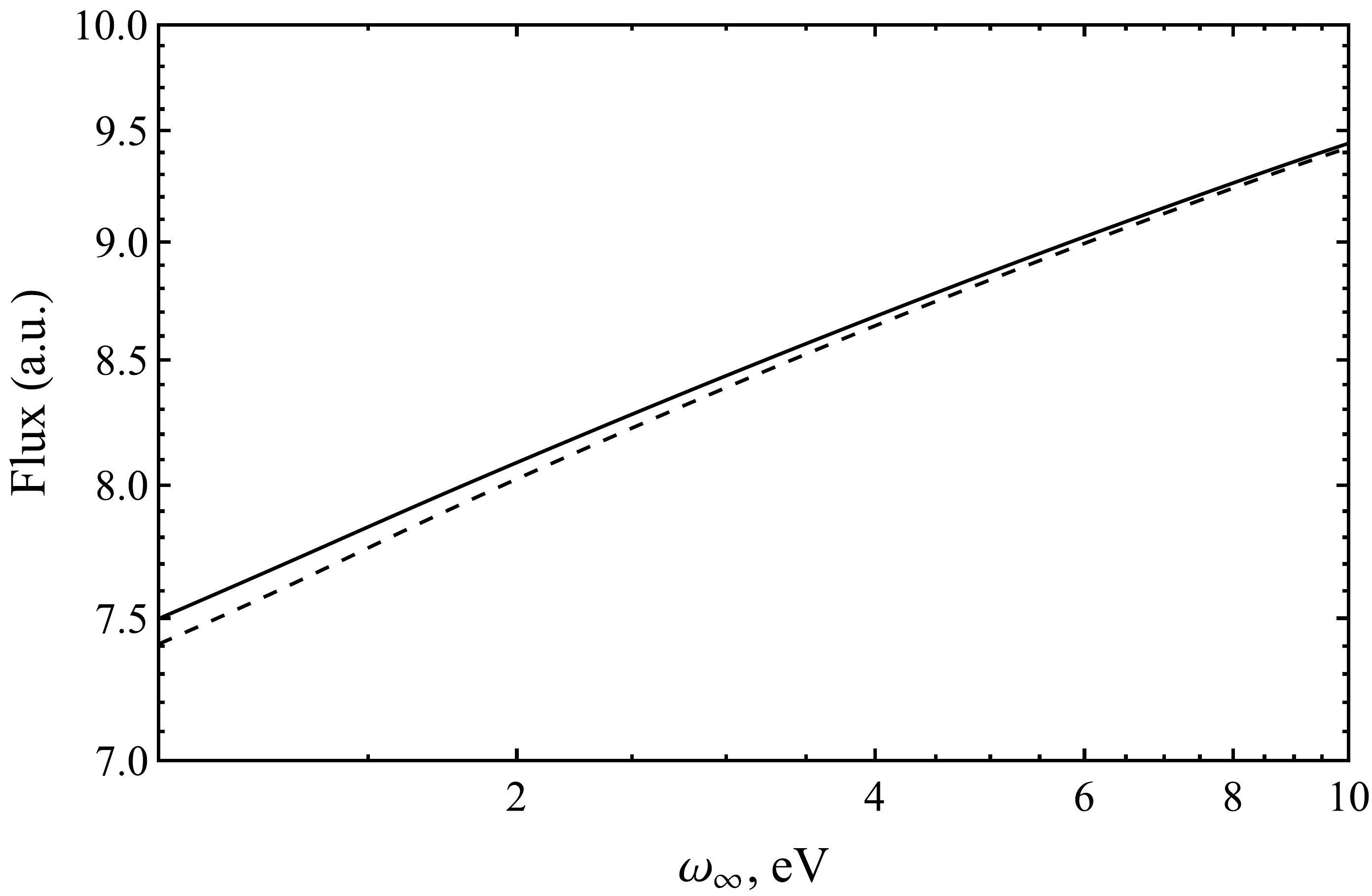}
\caption{Same as Figure \ref{fig:fig1} but for the free-ion model with $\left( \xi,\chi \right) = \left( 35^\circ, 5^\circ \right)$.
\label{fig:fig4}}
\end{figure}

In the case of a free-ion condensed surface, the emissivity in the O-mode is closer to that in the X-mode than in the fixed-ion limit. The optical spectra for $\left( \xi,\chi \right) = \left( 35^\circ, 5^\circ \right)$ are presented in Figure \ref{fig:fig4}; the intrinsic polarization degree is still negative, but closer to zero than in the fixed-ion case. The conversion decreases the total flux by around 12\%, which, again, can hardly be distinguished with observations. 

\begin{figure*}
\includegraphics[width=6cm]{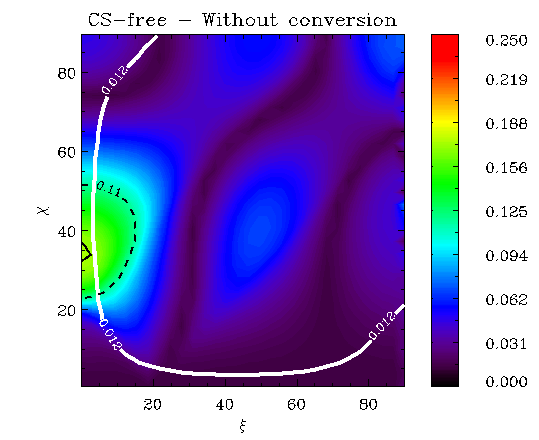}
\includegraphics[width=6cm]{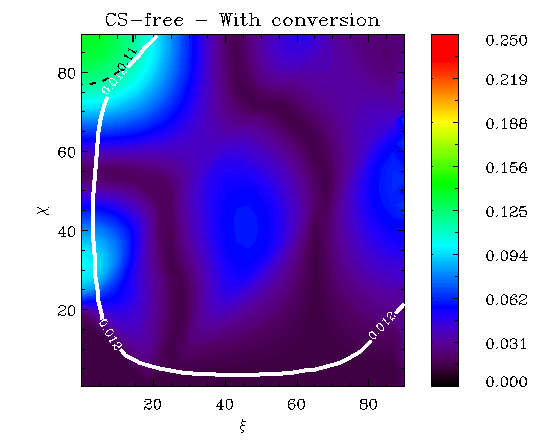}
\includegraphics[width=6cm]{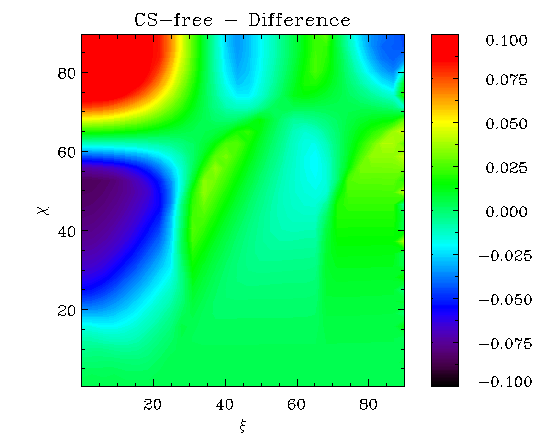}
\caption{Same as Figure \ref{fig:fig2} but for the free-ion model. 
\label{fig:fig5}}
\end{figure*}

\begin{figure*}
\centering
\includegraphics[width=7cm]{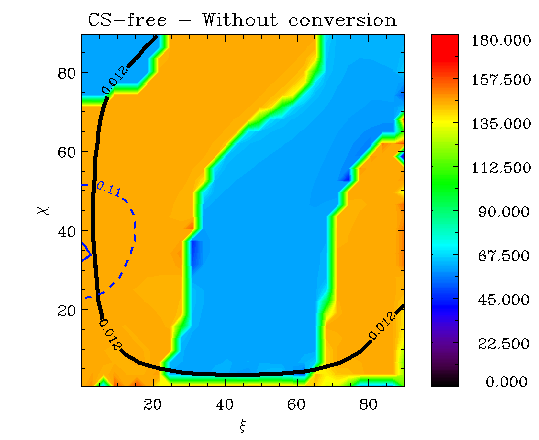}
\includegraphics[width=7cm]{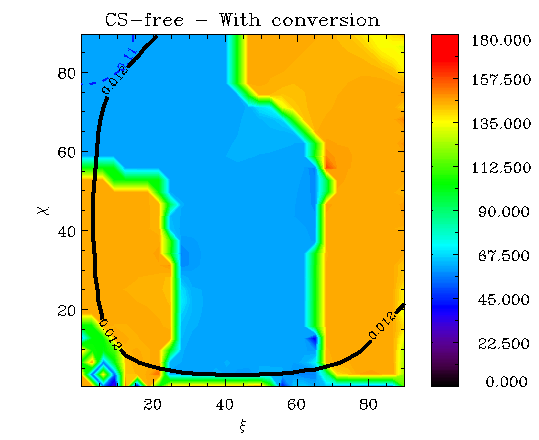}
\caption{Same as Figure \ref{fig:fig3} but for the free-ion model. The polarimeter frame was rotated by an angle $\hat{\chi}_p = 145.39^\circ$ with respect to the fixed frame.
\label{fig:fig6}}
\end{figure*}

The corresponding phase-averaged polarization fraction is illustrated in Figure \ref{fig:fig5}. As we can see, without conversion the free-ion model cannot simultaneously reproduce the measured value of 16.43\% and the pulsed fraction of 1.3\%. With 20\% of the O-mode flux being removed, $\Pi_{\mathrm{L}}$ can barely reach the $1\sigma$ error region. Also, note that at some geometries the difference in polarization fraction is positive, which implies that the modified flux consists primarily of the X-mode photons. This is better seen from our calculations of the polarization angle, shown in Figure \ref{fig:fig6}; the polarimeter frame is still rotated by $\beta=\hat{\chi}_p$ with respect to the fixed frame. Without conversion, $\chi_p$ is equal to $145.39^\circ$ in the $1\sigma$ region of $\hat{\Pi}_{\rm L}$, and changes to $55.39^\circ$ (i.e. $\hat{\chi}_p+90^\circ$) when the conversion is taken into account. This confirms confirms our conclusion that the flux becomes dominated by the X-mode photons. To match the observed value of $145.39^\circ$ in the top-left corner of the $\xi-\chi$ plane for the case with conversion, the polarimeter frame should be rotated by $235.39^\circ$ (or simply by $55.39^\circ$).

Finally, in the case of an atmosphere, the effect of conversion was negligible -- the total intensity decreased by less than 5\%. This is simply due to the fact that the ordinary photons make up about 7\% of the total flux for $\Pi_\text{L}^\text{EM} = 0.87$, and removing its fraction cannot produce observable results. For this reason, we do not show the plots of polarization fraction and angle, since the former changes by no more than 1\%.

\section{Discussion}

In this work we have calculated the effect of photon-axion mixing on thermal emission of RX J1856.5-3754 in the optical band, using realistic models of surface layers. The condensed surface in both fixed and free-ion limits turned out to be a promising target for detecting the change in phase-averaged polarization fraction, which can be significantly modified by a slight decrease in the O-mode flux. In the case of a gaseous atmosphere, the conversion had no measurable effects.

\begin{figure*}
\centering
\includegraphics[width=7cm]{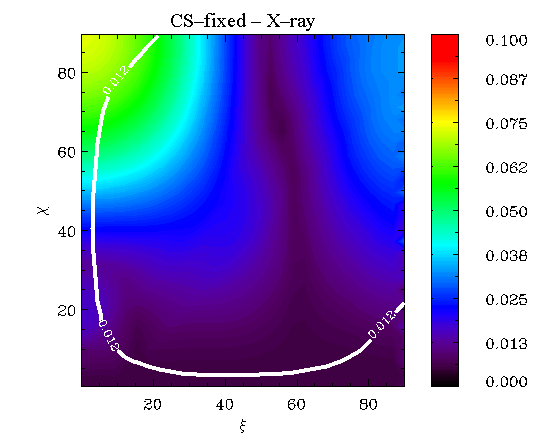}
\includegraphics[width=7cm]{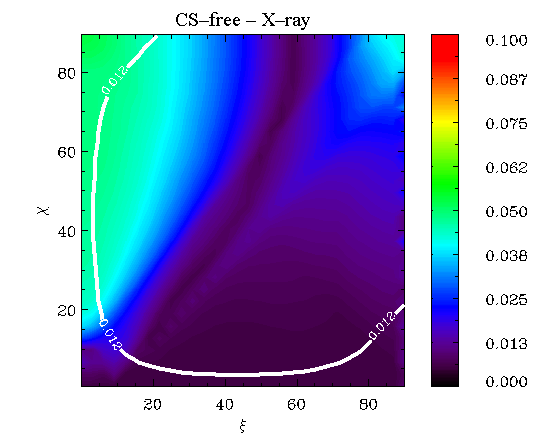}
\caption{Contour plots of the phase-averaged polarization fraction in the soft X-ray band (0.12 - 0.39 keV) for the fixed- and free-ion condensed surface, respectively. The photon-axion interaction has no effect at these energies.
\label{fig:fig_xray}}
\end{figure*}

Optical polarimetry alone is clearly insufficient to determine the origin of surface emission from RX J1856, since both an atmosphere and a condensed surface can reproduce the observed $\Pi_\text{L}$ and $\chi_p$ even if mixing is at work.
This degeneracy can be effectively removed using polarimetric observations at higher energies.
In fact, as pointed out in Paper II, the two models of stellar surface emission produce different polarization patterns in the soft X-ray band. For a condensed surface in both limits, $\Pi_{\mathrm{L}}$ is small for all possible source geometries, but non-negligible for some combinations of $\left( \xi, \chi \right)$, see Figure \ref{fig:fig_xray}. On the other hand, for an atmosphere $\Pi_{\mathrm{L}}$ is high for $\xi < \chi$, and low for $\xi \geq \chi$ due to the geometrical effects, see Section 4 and Paper II. However, for RX J1856, the latter case is ruled out because such geometries do not reproduce the measured $\hat{\Pi}_{\mathrm{L}}$ in the optical band; as a result, in the X-rays the polarization degree of an atmosphere should always be high. 

Since thermal high-energy photons cannot convert into axions (see Paper I), measuring the soft X-ray $\Pi_{\mathrm{L}}$ for RX J1856 can provide a simple method of constraining a range of axion parameters. If the polarization is significant, the emission is most likely atmospheric and the mixing will have no effect. On the other hand, if $\Pi_{\mathrm{L}}$ is low, the emission should originate from a condensed surface. In this case, the constraints can be obtained using the approach described in Sections 3 and 4. For a given set of $\left( m_a, g_{\gamma a} \right)$, we determine the converted fraction of optical O-mode photons and calculate the polarization fraction in free and fixed-ion limits. Then, we attempt to find a combination of $\left( \xi, \chi \right)$ which simultaneously reproduces the high and low-energy $\Pi_{\mathrm{L}}$, as well as the pulsed fraction of 1.3\%, taking into account the observational errors. If there will be no such geometry in both limits, we can infer that the considered values of $\left( m_a, g_{\gamma a} \right)$ lead to an unrealistic degree of conversion and can be excluded.

The effect of conversion on the optical polarization fraction at lower values of $g_{\gamma a}$ can be estimated as follows. As discussed in Paper II, the $\xi - \chi$ parameter space can be divided into two areas: $\xi < \chi$, where the observed $\Pi_\mathrm{L}$ is mainly determined by the intrinsic polarization at the source, and $\xi \geq \chi$, where the latter is distorted by the geometrical effects. Since for RX J1856 only the region $\xi < \chi$ is of interest, we can neglect the geometrical effects and calculate $\Pi_\mathrm{L}$ analytically using the intrinsic polarization degree, which is simply the relative difference between the fluxes in two modes (Equation \ref{eq:intrinsic}). As a result, the ratio $F_O/F_X = 1.39$ is required to match the measured $\hat{\Pi}_{\mathrm{L}}=16.43 \%$. The polarization is reduced to $\hat{\Pi}_{\mathrm{L}} - 1\sigma = 11.17 \%$ if the fluxes relate as $F_O/F_X = 1.11$, which can be achieved if 10\% of the O-mode photons convert into axions. For a surface magnetic field $B_p = 10^{13}$ G, this effect is reached at $g_{\gamma a} \geq 10^{-11}$ GeV$^{-1}$ and $m_a \leq 10^{-6}$ eV, see Paper I. Obviously, the actual exclusions will depend on the value of X-ray polarization fraction of RX J1856, and may be greater for other XDINSs with stronger magnetic fields \citep{2009ASSL..357..141T}.

Polarimetric observations are not the only method of determining the origin of thermal emission from XDINSs. Numerous studies have focused on the spectral properties of an atmosphere and a condensed surface to explain the well-known feature of the ``Magnificent Seven" -- their optical and ultraviolet counterparts exceed the extrapolation of X–ray blackbodies at low energies by a factor of 5 – 50 \citep{2011ApJ...736..117K}. Some works restrict to a single model of surface layers: a strongly magnetized atmosphere was suggested for RX J1605.3+3249 \citep{2007MNRAS.377..905M}, while the spectrum of RX J0720.4-3125 was attributed to the emission of a bare condensed surface \citep{2006A&A...459..175P}. However, recent studies prefer a combination of both, in which the atmosphere is relatively thin and the spectrum of outgoing radiation is also affected by the properties of metallic surface underneath it. Although the mechanisms of creation of such atmosphere are still debated, the model produced satisfactory results for RX J1308.6+2127 \citep{2010A&A...522A.111S, 2011A&A...534A..74H}, RX J1605.3-3249 \citep{2019A&A...623A..73P, Malacaria_2019}, RX J0720.4-3125 \citep{2017A&A...601A.108H}. 

In the case of RX J1856, X-ray data reveal different blackbody temperatures, $T^\infty_c \sim 40$ eV and $T^\infty_h \sim 60$ eV \citep{2012A&A...541A..66S}, as well as the absence of absorption lines or other spectral features. \cite{2007MNRAS.375..821H} were able to model the broadband spectrum with a thin hydrogen atmosphere above a condensed iron surface, although their magnetic field estimate $B_p \approx  4 \times 10^{12} \text{  G}$ somewhat contradicts the spin-down value $B_p \approx  1.5 \times 10^{13} \text{ G}$ obtained by \cite{2008ApJ...673L.163V}. Also, the model of \cite{2007MNRAS.375..821H} imposes a constraint on the magnetic field geometry, in addition to that of the X-ray pulsed fraction. The light curve can be explained if either $\xi$ or $\chi$ is small $\left(<6^{\circ}\right)$, while the other angle lies between $20^{\circ}$ and $45^{\circ}$.

In principle, our analysis can be generalized to the thin atmosphere as well, but implementing this model in the code of Paper II will require a state-of-the-art approach. Still, the effect of conversion can be estimated on the basis of the general properties of emission from a condensed surface covered by a thin atmosphere, see e.g. \cite{2012A&A...546A.121P}. At variance with a conventional atmosphere, optically thick from optical to X-ray energies, a thin atmosphere is designed in such a way to be thick at low energies and then become thin above $\approx 0.1$ keV because of the reduced free-free opacity with increasing frequency. As a result, the emergent radiation in the soft X-rays (and above) is that produced by the underlying condensate, while in the optical band it is that of a standard atmosphere. Since the radiation of the latter is predominantly polarized in the X-mode, we can infer that the mixing is unlikely to create observable signatures in the case of a thin atmosphere.

In our approach, the only missing component is the high-energy polarization data for RX J1856. Currently, a number of X-ray polarimetry missions are at an advanced stage of development.
The IXPE \citep{2016SPIE.9905E..17W}, a NASA Small Explorer (SMEX) mission expected to fly at the end of 2021, will measure polarization in the $2-8$ keV energy range.
Even though the IXPE bandpass is at higher energies than those where the emission from XDINSs peaks, the latter can be reached with its follow-up mission, the X-ray Polarization Probe \citep{2019arXiv190710190J}, which promises a sensitivity improvement by a factor of $3-10$ over IXPE and an energy bandpass broadened from $2-8$ keV to $0.1 - 60$ keV.
Other promising missions include eXTP \citep{2019SCPMA..6229502Z} and PiSoX \citep{2020SPIE11444E..2YM}.

Apart from the phase-averaged optical and X-ray polarimetry, no other diagnostics seems to be as useful for determining the physical state of stellar surface layers.
Phase-resolved polarimetry, on the one hand, can provide information on the phase dependence of $\Pi_\mathrm{L}$ and $\chi_p$, which is particularly sensitive to the source geometry and surface emission model \citep{2015MNRAS.454.3254T,2020MNRAS.492.5057T}.
This makes it more descriptive than phase-averaged measurements, which are largely degenerate with respect to the angles $\left( \xi,\chi \right)$ since $\chi_p$ remains constant in large regions of the $\xi-\chi$ plane (see Figures \ref{fig:fig3} and \ref{fig:fig6}).
However, in the optical band, such observations are hardly possible due to the faintness of the optical counterpart of RX J1856 ($V\sim 25.5$), even with next-generation telescopes such as the European Extremely Large Telescope (ESO-ELT\footnote{See \tt{https://elt.eso.org/}.}).
In the X-rays, it is unclear whether the high-energy tail of RX J1856 will be bright enough for phase-resolved measurements to be performed by upcoming soft X-ray polarimetry missions with affordable exposure times.
Spectral fitting, on the other hand, is even more degenerate than phase-averaged polarimetry, since it generally yields too many sets of atmospheric and condensed surface models which can satisfactorily reproduce the observed broadband spectra \cite[see Paper II and][]{2014PhyU...57..735P}.

Our study can be extended to other XDINSs as well, since their low surface temperatures and strong magnetic fields are sufficient both for the surface layers to enter a condensation phase and for the mixing to have a significant effect on the optical emission. 
However, the low-energy counterparts of the remaining XDINSs are much fainter than that of RX J1856 \citep{2011ApJ...736..117K}, and polarization measurements are beyond the capabilities of present generation telescopes, like the VLT. 
Only the sensitivity improvement promised by next-generation instruments such as ESO-ELT will allow to extend our study to more XDINSs.
Other classes of neutron stars (e.g. magnetars) do not appear as suitable, since their optical emission, the origin of which is still uncertain, is likely due to non-thermal processes in the magnetosphere \citep[see][for a review]{2015RPPh...78k6901T}, and the contribution from thermal surface emission is too faint for mixing to produce an observable effect.

\section*{Acknowledgements}

We thank Sergei Popov and Sergey Troitsky for numerous helpful comments. The work of RT and RT 
is partially supported by the Italian MUR through grant UNIAM (PRIN 2017LJ39LM).
\bibliographystyle{aasjournal} 
\bibliography{bibliography}

\end{document}